\documentclass[
	aps, prd,onecolumn,reprint
	floats, floatfix,
	amsmath, amssymb, amsfonts,
	superscriptaddress, showpacs, footinbib
]{revtex4-1}

\usepackage{epsfig} 
\usepackage[hang,nooneline]{subfigure}
\usepackage{amsmath}
\usepackage{amsfonts}
\usepackage{graphicx}

 \usepackage{amssymb}

\date{\today} 

\newcommand{\be}{\begin{equation}}
\newcommand{\ee}{\end{equation}}
\newcommand{\bea}{\begin{eqnarray}}
\newcommand{\eea}{\end{eqnarray}}
\newcommand{\bml}{\begin{mathletters}}
\newcommand{\eml}{\end{mathletters}}

\begin{document}






\title{Excited DeSitter brane worlds localized by a  kink.}
\renewcommand{\thefootnote}{\fnsymbol{footnote}}
\author{Y. Brihaye\footnote{yves.brihaye@umons.ac.be}}
\affiliation{Facult\'e des Sciences, Universit\'e de Mons, 7000 Mons, Belgium}
\author{T. Delsate\footnote{terence.delsate@umons.ac.be}}
\affiliation{Facult\'e des Sciences, Universit\'e de Mons, 7000 Mons, Belgium}
\affiliation{CENTRA, Instituto Superior T\'ecnico, Avenida Rovisco Pais 1, 1049-001 Lisboa, Portugal, EU}

\date{\today}
\setlength{\footnotesep}{0.5\footnotesep}

\begin{abstract}
We reconsider, in five-dimensional space-time, the issue of thick brane 
localized in the extra dimension by a kink formed by a scalar field.
The localization is achieved by a sine-Gordon potential. 
Apart from a fundamental brane (discovered by Koley and Kar \cite{koley_kar})
where the scalar field is a monotonic function of the extra dimension),
we show that a series of new solutions exist as well, labelled by the number of zeros of the scalar
field.  These solutions are regular, localized on the brane and mirror symmetric with respect to the
extra dimension. They form a tower of "excited branes".
The study of some perturbations of the solutions
reveals that the new solutions are not stable. 
Finally, fermions are  coupled to the scalar field by means of a Yukawa potential
and their localization in the background of the new solutions is examined. 
It turns out that the excited branes can localize left and right chiral fermion
either on the brane and/or in the bulk but close to the brane. 
\end{abstract}

 \maketitle
\section{Introduction}
One of the main problem in the theoretical description of the Universe
is the cosmological constant problem \cite{ccp}. One of the most promising attemp to solve it is based on the "brane-world"
models \cite{dvali}; the basic idea is that our Universe  
is represented
by a 3+1-dimensional subspace (a three-brane) embedded into an higher 
dimensional space-time (the bulk). In this context, 
the first challenge is to reconciliate the extra dimensions 
with the fact that low energy physics is very well described by a conventional 
3+1-dimensional
space-time in which numerous laws (e.g. Newton's law) are tested with a very
high degree of accuracy.   It is nevertheless challenging to emphasize the
Einstein equations in higher dimensional space-times and to study and classify
the possibly relevant solutions. 
One way to attack the various problems consists in trying to
localize the matter fields and gravity on the brane in the case where the extra dimensions
are non compact \cite{rs,shapo}; another way would be to have solutions
where the geometry of the extra-dimensions naturally comes out to be 
a compact manifold \cite{rs}.
\par Here we will follow the ideas developped in several papers 
\cite{cho,chovil,japanese} and consider branes which are supported
by appropriate topological deffects living in the extra-dimensions.
In these papers, the brane is supposed to be Minskowski space-time
(no vacuum energy) or de Sitter space-time with a cosmological constant
(or brane tension) leading to an inflating brane. No  bulk cosmological
constant is assumed. In contrast, in \cite{shapo} the authors take advantage
of a cosmological constant in the bulk : $\Lambda_{4+n} \neq 0$ and manage
to enforce the localisation of gravity by a fine tuning of $\Lambda_{4+n}$
in function of the other constants characterizing the topological defects.

Assuming only one extra dimension, a lot of work has been done about thick brane
scenarios \cite{df,ce,g,c} where five-dimensional gravity is coupled to one or more scalars and
an appropriate potential leading to localized solutions in the extra dimension.
The relation of gravity localisation with different types of potential for 
one scalar field was analyzed namely in \cite{bg} and in \cite{deformed_1,deformed_2,deformed_3};
here the relevant potential derive from a superpotential allowing for explicit solutions to the 
field equations. The Sine-Gordon soliton potential was used in \cite{koley_kar}.
In a series of papers, two scalar fields were used to construct a so called Bloch brane \cite{bg2,bnrt,acfg}
which is also based on a superpotential.

Here, we will reconsider the existence of branes within the sine-Gordon model. 
One feature of the brane solution found in \cite{koley_kar} is that it exists only for a particular
relation between the five dimensional Newton constant  and the bulk cosmological constant $\Lambda$. 
In this paper, we reinvestigate the model of \cite{koley_kar}  and study the field equations
numerically for generic value of the constant $\Lambda$. 
Our numerical result strongly suggest that the Koley-Kar brane-world is the first solution 
of a tower of solutions characterized by the number of nodes of  the scalar field. All these solutions
are mirror symmetric with respect to the four dimensional brane and exist for specific values
of the cosmological constant (like a spectral value while the five dimensional Newton constant is fixed. 
For the other values of $\Lambda$, the warp factor become singular on one side of the bulk space.
The excited solutions have important consequences on the localization of masless fermion on the brane.
In particular, both chiral component of the fermion can be localized on or close to the brane while it is know that
only the left component is localizable in the background of the Koley-Kar fundamental solution.
We also checked that the tower-like  solutions of the type studied here are also present with other
type of potential admitting kink in the flat space, e.g. the so called $\lambda \phi^4$ potential.
The model and the equations are presented in Sect. II. In particular we consider the case where the
four dimensional brane is an Einstein space with a independant cosmological constant.
  The various type of solutions are discussed 
in Sect. III and Sect. IV is devoted to the localization of the fermions. Finalize, we analyze the stability
property of the first few elements of the tower in Sect. V and give some conclusions in Sect. VI.

\section{The model and the equations}
Along with Koley and Kar \cite{koley_kar} (and more recently \cite{liu}), we consider the action
\begin{equation}
\label{action}
  S = \int d^5 x \sqrt{-g}[\frac{1}{2 \kappa_5^2}(R- 2 \Lambda) - \frac{1}{2} g^{MN} \partial_M \phi \partial_N \phi - V(\phi)]
\end{equation}
where $\kappa_5^2= 8 \pi G_5$ with $G_5$ the 5-dimensional Newton constant and $\Lambda$ is the 5-dimensional
cosmological constant. The real scalar field $\phi$ interact through the potential $V(\phi)$.
In this note we have  considered both the  sine-Gordon potential and the  quartic double  potential 
\begin{equation}
\label{potential}
 V(\phi) = p (1+\cos(\frac{2 \phi}{q})) \ \ \ {\rm or} \ \ \ V(\phi) = p (\frac{\phi^2}{q} - 1)^2
 \end{equation}
where $p,q$ are real constants. These potentials lead qualitatively to the same results but we will discuss
essentially the case of the sine-Gordon potential.
We are interested in solutions of the field equations associated to the action (\ref{action}) with a metric of the form
\be
\label{metric}
ds^2 = e^{-2A(y)} (ds_4^2) + dy^2
\ee
where the extra coordinate is denoted $y$ and the function $\exp({-2A(y))}$ is the warp factor 
(note that the function  $A(y)$ in our notation differs by a sign from the one of \cite{liu}).
The scalar field is assumed to depend on $y$ only~: $\phi = f(y)$. 
The 4-dimensional space whose metric is denoted $ds_4^2$ and is assumed to be an Einstein space, i.e. such that
\be
R^{(4)}_{\mu \nu} =  \frac{H^2}{4} g^{(4)}_{\mu \nu} 
\ee
where $H^2$ is a constant which can be negative. A real (resp. negative, nul) $H^2$ corresponds to deSitter 
(resp. Anti de Sitter, Minkowski) 4-dimensional space-time but the metric $g^{(4)}$ can actually describe \emph{any} Einstein space.
 
The relevant components of the 5-dimensional Ricci tensor are then
\be
            R_0^0 = R_1^2 = R_2^2 = R_3^3 = 3 e^{2A} H^2 + A'' - 4 (A')^2  \ \ , \ \ R_4^4 = 4 A'' - 4 (A')^2
\ee
from now on, the prime denotes the derivative with respect to $y$.
The field equations read
\begin{eqnarray}
       A''  &=& \frac{\kappa}{3} (f')^2 + H^2 \exp {2 A}  \ \   , \ \ \kappa \equiv \kappa^2_5 \label{eqa} \\
       A'^2 &=& \frac{\kappa}{12} (f')^2 - \frac{1}{6} \kappa V(f) - \frac{\Lambda}{6} + H^2 \exp {2 A} \label{eqc} \\
       f''  &=& 4 A' f' + \frac{\partial V}{\partial f} \label{eqf}
\end{eqnarray}
These equations depend in principle on five constants~: the parameters of the potential $p,q$
the gravity parameter $\kappa$ and the bulk and the brane cosmological constants, respectively $\Lambda, H$. 
The parameter $q$ defines the scale of the scalar field and can
be set to one. Independently, $p$ can be set to a fixed value by a suitable rescaling of $y, \Lambda$ and of $\kappa$.
In the following we adopt the normalisation  $q=1$, $p=1/4$.  Using (\ref{eqa}) and (\ref{eqf}) to determine $A(y),f(y)$,
the cosmological constant $\Lambda$ then appears as a "constant of motion" through Eq.(\ref{eqc}). 
As a consequence, only $\kappa, H$ need to be specified to construct a solution.

\section{Analysis of the solutions}
\subsection{Case $H=0$}
The equations (\ref{eqa}),(\ref{eqf}) have to be solved with appropriate boundary conditions. 
Without loosing generality, the coordinate $y$ can be translated in such a way that $\phi(0)=0$;
we want the sine-Gordon field to approach a vacuum for $y \to \infty$, so we require $\phi(\infty)=\pi/2$.
In addition, the condition $A(0)=0$ can be set by a suitable normaliation of $t$ 
, the brane is then located  at $y=0$. 
This leaves the parameter $A'(0) \equiv C$ to be chosen  in order to 
specify a  boundary value problem.

The hypothesis of a symmetric space time under the reflection $y \to -y$ is fulfilled if $C=0$. 
 However, assuming for a while space-time limited to the subspace with 
 $y\geq 0$, the equations can be integrated for $y \in [0, \infty]$ with  arbitrary
 values of $C$. The  boundary conditions are then
\be
             A(0) = 0 \ \ , \ \ A'(0) = C \ \ , \ \ f(0) = 0 \ \ , \ \ f(\infty) = \frac{\pi}{2}
\ee 
For $C=0$, the explicit solution of Koley, Kar is recovered
\be
        A(y) = \frac{\kappa}{3} \ln \cosh(ky) \ \ , \ \ f(y) = 2 \arctan(\exp ky) - \frac{\pi}{2}
\ee 
with
\be
        k^2 = \frac{12 p}{4 \kappa + 3} \ \ , \ \ \Lambda = - \frac{8 p \kappa^2}{4 \kappa + 3}
\ee
In particular, the bulk cosmological constant is fixed by the five dimensional coupling constant and the 
parameter $p$. 

\begin{figure}
\centering
\epsfysize=8cm
\mbox{\epsffile{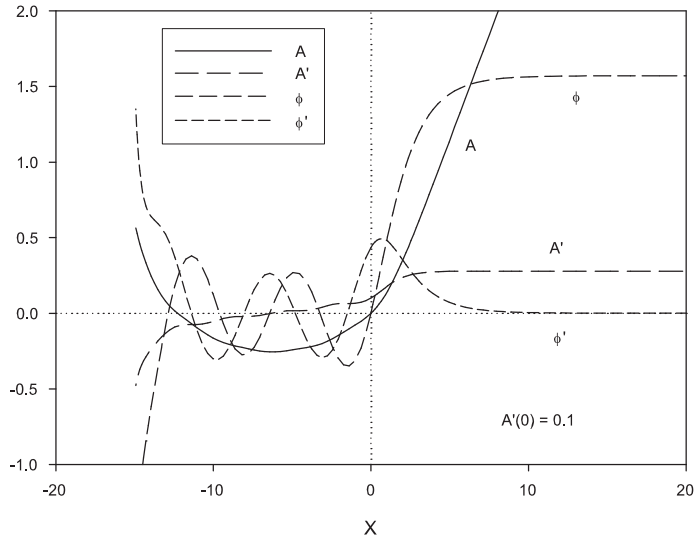}}
\caption{\label{fig_2_b}
The functions $A,A',f,f'$ for $\kappa=1$, $p=0.25$ and $C=0.1$}
\end{figure}

For $C > 0$, the system of equation above has, to our knowledge, no explicit solution. We therefore
solve the equation by a numerical method. 
We used a collocation method for boundary-value ordinary
differential equations, equipped with an adaptive mesh selection procedure \cite{colsys}.
Our solutions were constructed with a relative error of order $10^{-8}$.
The  solution corresponding to $\kappa_5=1$ and $C=0.1$ is presented on Fig. \ref{fig_2_b}.  
Continuing the numerical solution on the negative part of the $y$-axis,
the numerical results show  that, for generic values of $C$, the solution 
develops oscillations and then becomes divergent when $y$ is small enough. 
Accordingly, the solution cannot be continued of the full axis $y \in [-\infty, \infty]$.
The  analysis reveals, however, that for fine tuned values of the 
parameter $C$, the solution has  $f(y_c)= A'(y_c)=0$ for some $y_c$. 
For these
values of $C$ and $y_c$, the solution can be continued into a regular, mirror symmetric solution
(after a translation by $y_c$)  on the full axis $y$. 

\begin{figure}[ht]
  \begin{center}
    \subfigure[$A,f$]{\label{soliton_a0_e2}\includegraphics[scale=0.55]{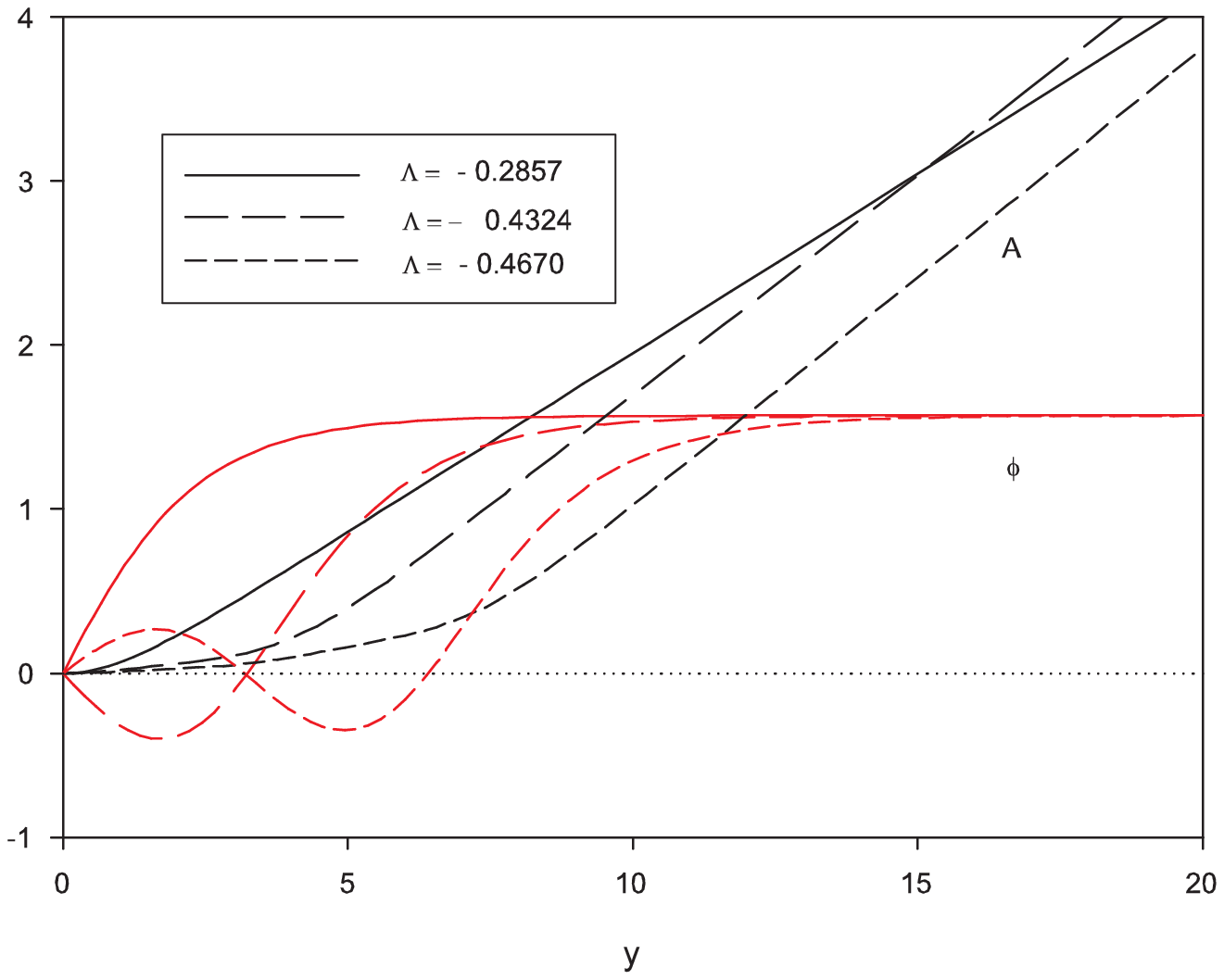}}
    \subfigure[$A', f'$]{\label{soliton_mpsi_e2}\includegraphics[scale=0.55]{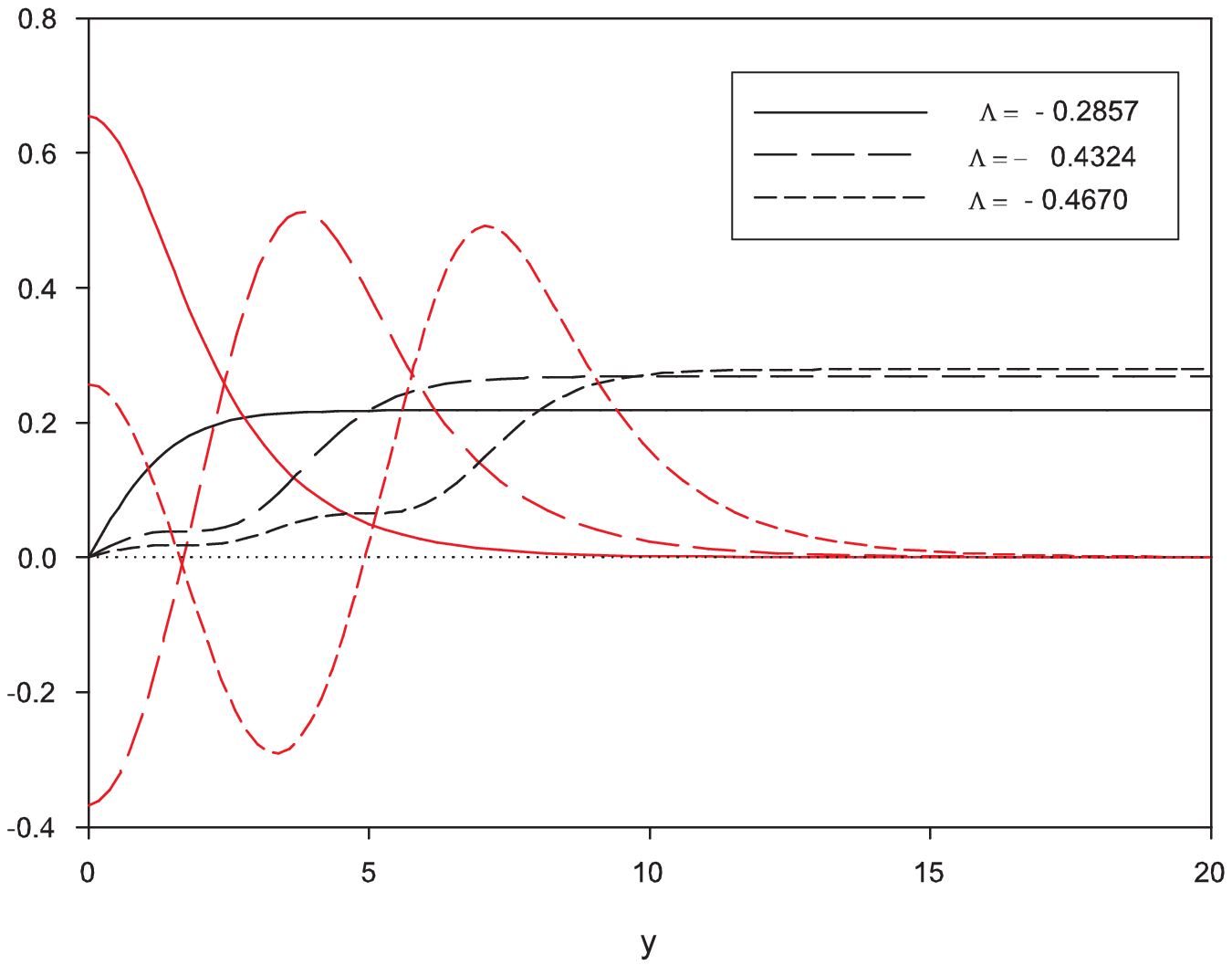}}
    \end{center}
   \caption{The metric function $A$ and scalar function $f$ for the  solutions with one, two and three nodes(left),
the corresponding derivatives (right)
         }
\label{fig_N_123} 
 \end{figure}

We believe that solutions with an arbitrary
number of oscillations in the core exist, forming an infinite tower labelled by $N$,
the number of nodes of $f(y)$ on $y \in [0,\infty]$ (corresponding to a total on $2N-1$ nodes).
We constructed the first three elements of this tower; 
the profiles  are presented  on Fig. \ref{fig_N_123} ($A,f$ on the left part,
 $A',f'$ on the right part).
Several parameters characterizing the brane solutions, namely the bulk cosmological constant
and the value $|f'(0)|$ are represented as functions of $\kappa_5$ on Fig. \ref{fig_N_123}
for the
first three solutions. The curves corresponding to $N=1$ is known analytically;
the $A$ function is such that $A'(\infty)= \sqrt{-\Lambda/6}$.
\begin{figure}
\centering
\epsfysize=8cm
\mbox{\epsffile{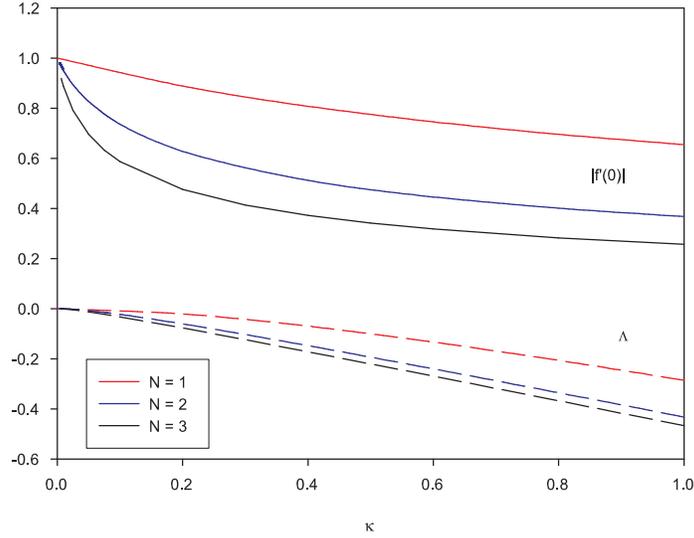}}
\caption{\label{data_N_123}
The parameters $|f'(0)|$ and $\Lambda$ as functions of $\kappa_5$ for $N=1,2,3$}
\end{figure}

The pattern of solutions of the sine-Gordon equation (i.e. in the probe limit $\kappa_5 = 0$)
suggests that the new solutions, found for $\kappa > 0$, are nothing else than  the first elements of a tower 
of gravitating solutions labelled by the number of nodes of the scalar function. We now present  arguments 
supporting this statement. With our conventions ($q=1,p=1/4$ and assuming the kink to be centered at the origin: $f(0)=0$), 
the sine-Gordon kink has $f'(0)=1$, $f(\infty)= \pi/2$. 
For the following, let us denote a "half-kink" the part of this solution between $y=0$ and $y = \infty$.
Integrating the sine-Gordon 
equation with $f(0)=0, f'(0)=s$ with $s < 1$ leads to  oscillating solutions; these are periodic
functions
whose zeros (which we note $y=y_n$) are such that $f'(y_n)=\pm s$. 
Accordingly, configurations can be
constructed by "gluing" a half-kink (suitably translated) at one of the zeros 
of a periodic solution corresponding to a value of $s$ arbitrarily close to $s=1$. 
Such configurations 
are off course {\it not} differentiable  at $y = y_n$ but the coupling to gravity somehow regularizes them
and leads to the family of regular solutions discussed above. Our numerical results strongly confirms this
interpretation. 

Replacing the sine-Gordon potential by the double well quartic potential given in (\ref{potential}) 
leads to the same pattern.
The occurence of the  "quasi-solution" configurations in approaching the probe limit  makes, however, the numerical
analysis rather involved for $\kappa_5 << 1$ for the two potentials.

\subsection{Case $H^2<0$}
Let us pose $H^2 = - K^2$.
In this case, we could not find explicit solutions for generic values of $\kappa$. 
For $\kappa_5 = 0$,
the equation for $A$ can be solved, leading to
\be
       A(y) = \frac{1}{2} \log (1 - {\rm tanh}^2(Hy) ) \ \  \longrightarrow \exp(-2 A) = {\rm cosh}^2(Hy) \ \ .
\ee
This function is such that $A  = - K^2y^2/2$ for $y << 1$ and  $A = - |K| y$  for $y >> 1$.
Accordingly, the warp factor $\exp(-2A)$ increases for large $y$ and the brane cannot be localized.
Solving the equations numerically for $\kappa_5 > 0$, reveals that, asymptotically, the warp factor
keeps diverging for large $y$. Several plots of such solutions are presented in Fig. 4.
\begin{figure}[ht]
  \begin{center}
    \subfigure[$A,f$]{\label{soliton_a0_e2_b}\includegraphics[scale=0.55]{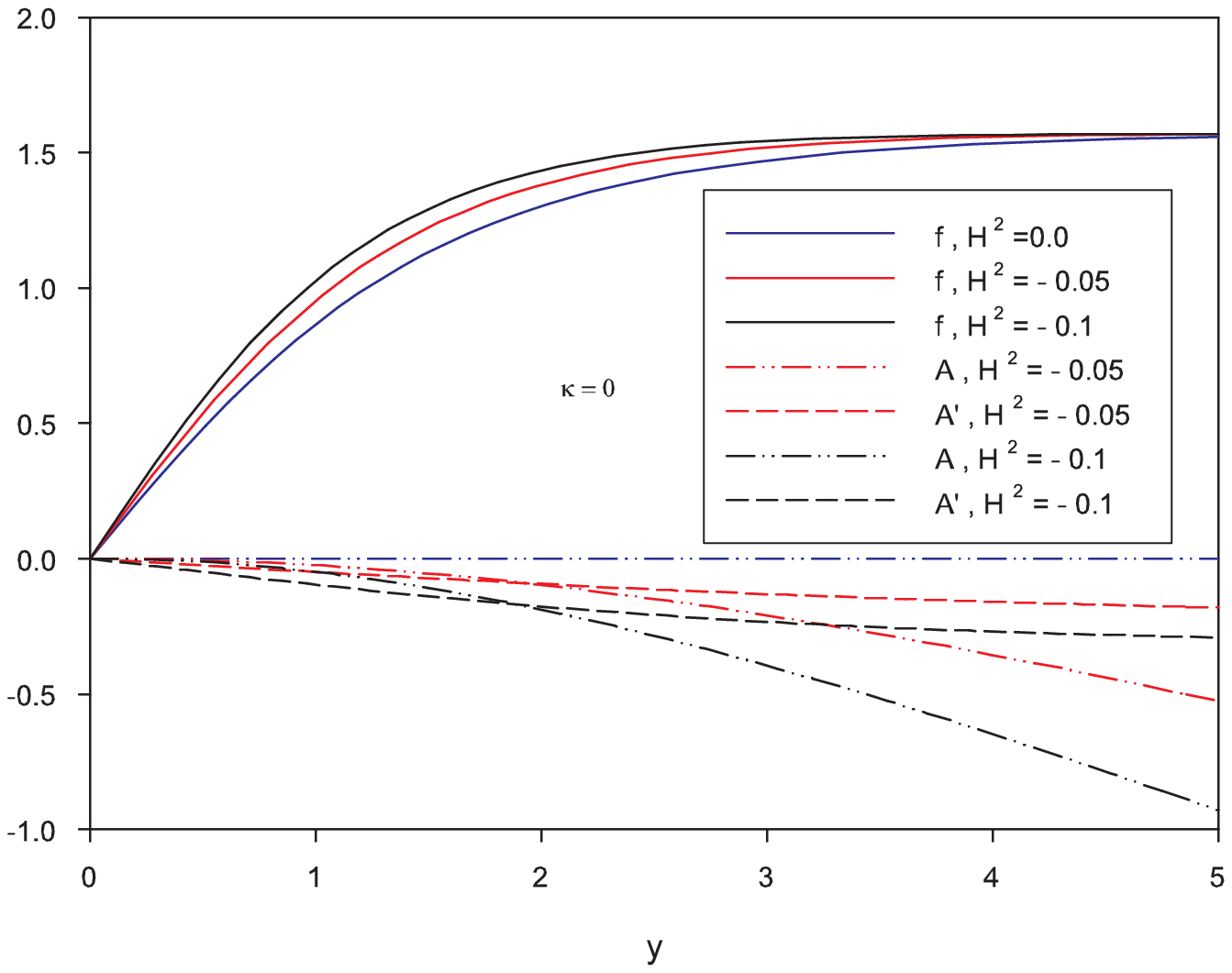}}
    \subfigure[$A,f$]{\label{soliton_mpsi_e2_b}\includegraphics[scale=0.55]{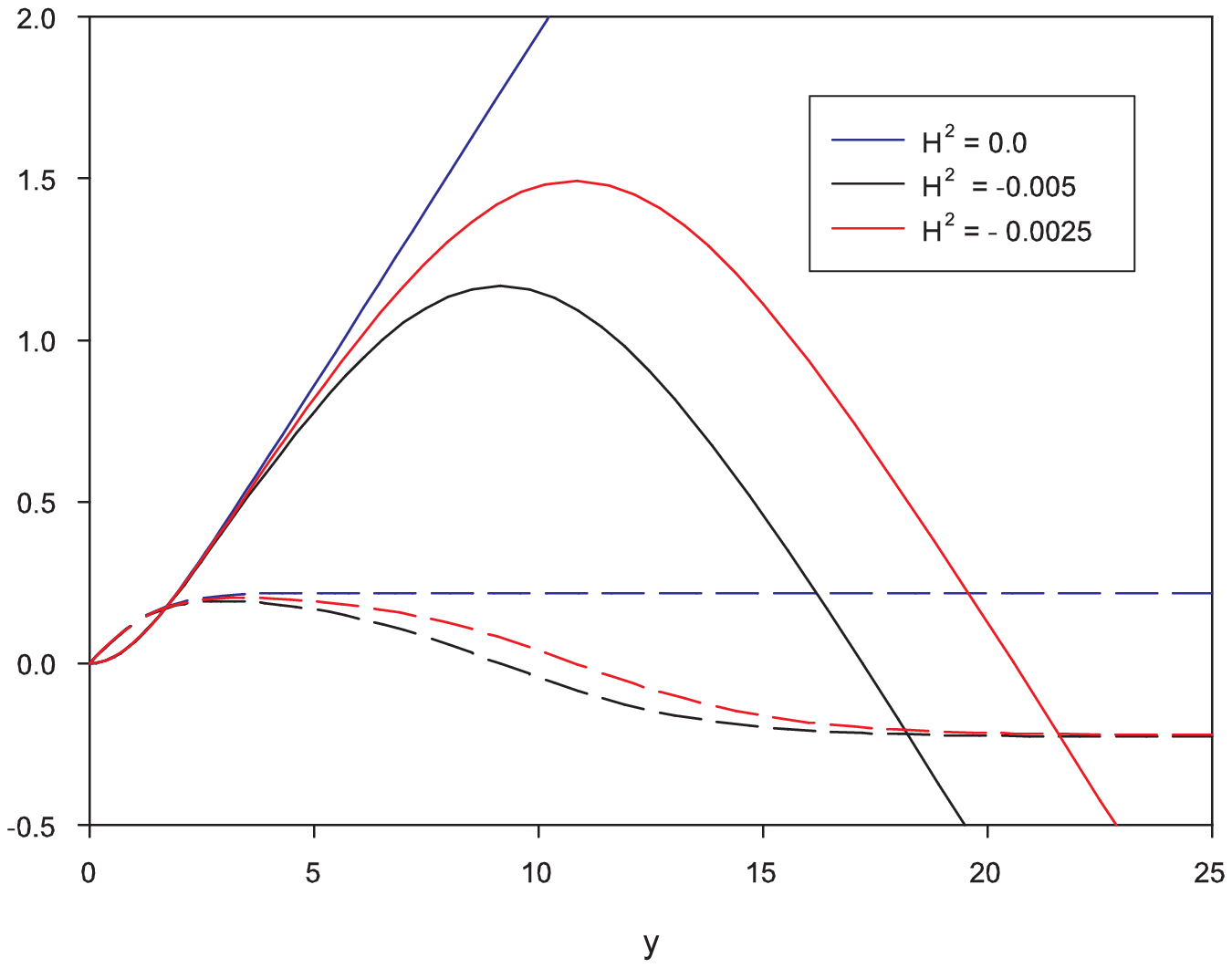}}
    \end{center}
   \caption{Left: The metric function $A$ and scalar function $f,f'$ for $\kappa=0$ and three values of $H^2$.
   Right: the metric function $A$ (solid) and the derivative (dashed) for $\kappa =1$ and three values of $H^2$.}
\label{fig_cc_123} 
 \end{figure}

\subsection{Case $H^2>0$}
In this case also, we could not find explicit solutions for generic values of $\kappa$. For $\kappa = 0$,
the equation for $A$ can be solved, leading to
\be
       A(y) = \frac{1}{2} \log (1 + {\rm tan}^2(Hy) ) \ \  \longrightarrow \exp(-2 A) = {\rm cos}^2(Hy)
\ee
It is such that $A(y) = H^2 y^2/2 + O(y^3)$ for $y << 1$. The warp factor vanishes  
 $y  = \pi/(2H)$, while the components of the Riemann tensor remain finite. 
 Solving (numerically) the equation for $f(y)$ in this background
reveals that the scalar field does not stabilize to a constant for $y \to \pi/(2H)$. 

Investigating  the equations numerically for $\kappa > 0$ reveals that  the peculiar behaviour of the $\kappa=0$
limit (in particular the vanishing of the warp factor at a finite value extra dimension coordinate $y$) persists.
\section{Localization of the fermions on the brane}
The coupling of the Dirac action to the brane is usually realized by means of the fermionic action
\be
      S_F = \int d^5 x \sqrt{-g} ( \overline \psi \Gamma^M D_M \psi - 
      m \overline \psi \psi - \eta \overline \psi F(\phi) \psi \ .
\ee
The  parameter $\eta$ represents the coupling of the fermion to the scalar field. The function $F(\phi)$ is usually
chosen as $F(\phi)= \phi$ leading the a conventional Yukawa potential.
(the alternative choice  $F(\phi)= \sin(\phi)$ is also used \cite{koley_kar,liu}). 
The matrices $\Gamma^M$ are the Dirac matrices in curved space.
To solve the underlying Dirac equations, one usually adopts a coordinate $z$ defined according to
$dz = exp(A(y)) dy$. This transforms the line element into a conformally flat metric
\be
     ds^2 = e^{-2A(z)} (\eta_{\mu \nu} dx^{\mu} dx^{\nu} + dz^2)
\ee
With this coordinate, the Dirac matrices are given by $\Gamma^M = (e^{A}\gamma^{\mu},e^{A}\gamma^5 )$ 
where  $\gamma^{\mu}, \gamma^5$ are the usual Dirac matrices in flat space.
The fermion function is usually decomposed according to
\be
\label{fermion}
\psi(x,z) = \sum_n \psi_{L,n} \alpha_{Ln}(z) + \sum_n \psi_{R,n} \alpha_{Rn}(z)
\ee
with the chiral spinor $\psi_{L,n} = - \gamma^5 \psi_{L,n}$ and $\psi_{R,n} =  \gamma^5 \psi_{R,n}$
and the sum $n$ run over the spectrum of the 4-dimensional solutions.

After some standard manipulations (see e.g. in \cite{koley_kar},\cite{liu}), the effective potential
determining the $z$-dependent prefactors $\alpha_{Ln}(z)$,  $\alpha_{Ln}(z)$ 
for the  chiral components of the Dirac spinors are given by
\be
     V_L(y) = e^{-2A} (\eta^2 \phi^2 - \eta \frac{\partial \phi}{\partial y} + \eta \phi \frac{\partial A}{\partial y})|_{y=y(z)}
\ee
\be
     V_R(y) = e^{-2A} (\eta^2 \phi^2 + \eta \frac{\partial \phi}{\partial y} - \eta \phi \frac{\partial A}{\partial y})|_{y=y(z)}
\ee
It is known that the fundamental Koley-Kar brane leads to a vulcano shaped potential $V_L$ which 
 can localize the fermion on the brane while $V_R$ is bell shaped,
admitting no bound states. Accordingly,
the right handed fermions cannot  be localized on the brane.
\begin{figure}
\centering
\epsfysize=8cm
\mbox{\epsffile{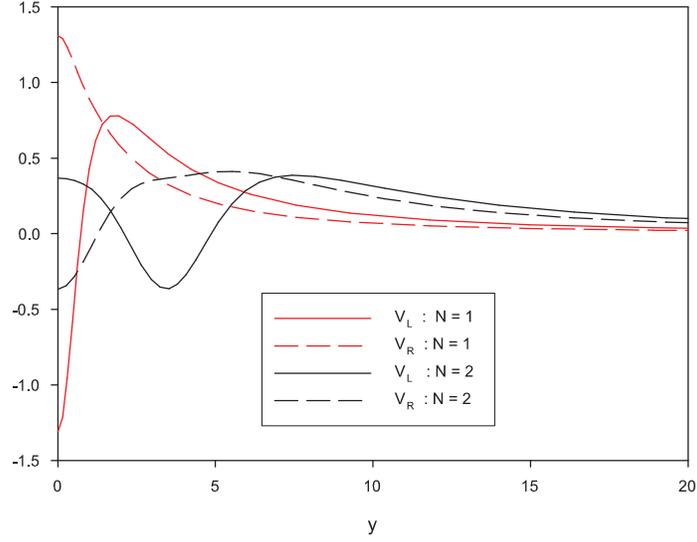}}
\caption{\label{fig:potential}
The effecive potentials $V_L$ and $V_R$ for $\kappa=1$, $\eta = 1$ for the fundamental (red) and first exited (black) brane}
\end{figure}

Owing on the new excited branes constructed in the previous section, it is natural to study
the shape of $V_L$ and $V_R$ corresponding to the excited solution. This is illustrated by Fig. \ref{fig:potential}.
It turns out that the shape of the potential corresponding to the exited brane 
is quite different from the one of the fundamental brane and presents some interesting features.
On Fig.\ref{fig:potential} the  potentials $V_{L,R}$ corresponding to $\kappa=1$, $\eta=1$ and the N=1 and N=2 
solutions are superposed. 
For the first excited brane, it turns out that $V_R$ presents a negative valley on the brane (i.e. for $z=0$) and 
at least one bound state should exist there. 
In contrast, the potential $V_L$ presents a local maximum on the brane
and a local minimum at some finite value $y = \pm y_m$. It means  that 
left handed fermions could be bounded in this region; the corresponding fundamental wave function should be 
localized symmetrically in the two valleys.
Decreasing the Yukawa coupling constant $\kappa$, it turns out that 
the  potentials $V_L$ and $V_R$ both present a local minimum on the brane; accordingly  
both species of fermions could be localized. 

\begin{figure}[ht]
  \begin{center}
    \subfigure[$V_{LR}$]{\label{potential1}\includegraphics[scale=0.55]{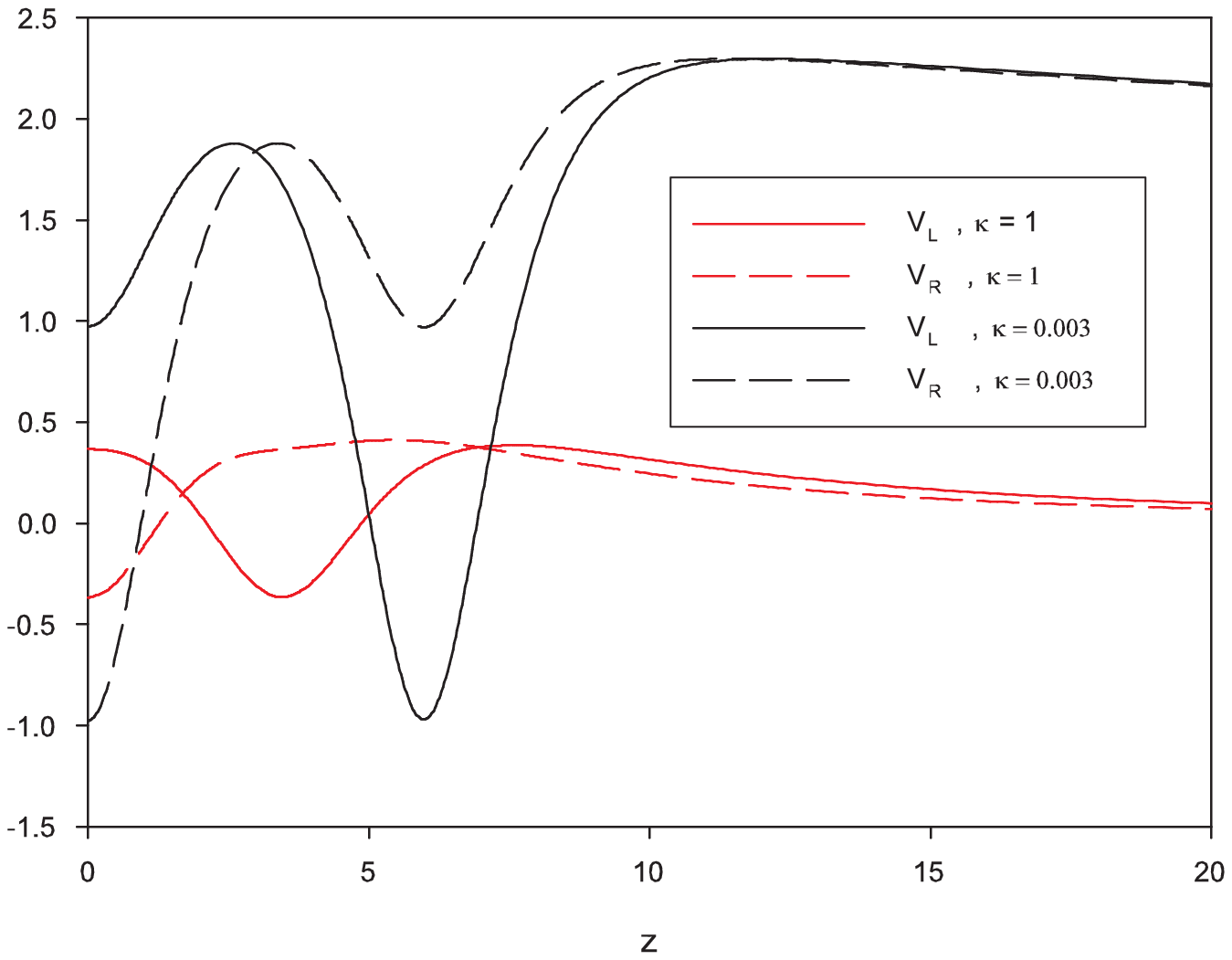}}
    \subfigure[$V_{LR}$]{\label{potential2}\includegraphics[scale=0.55]{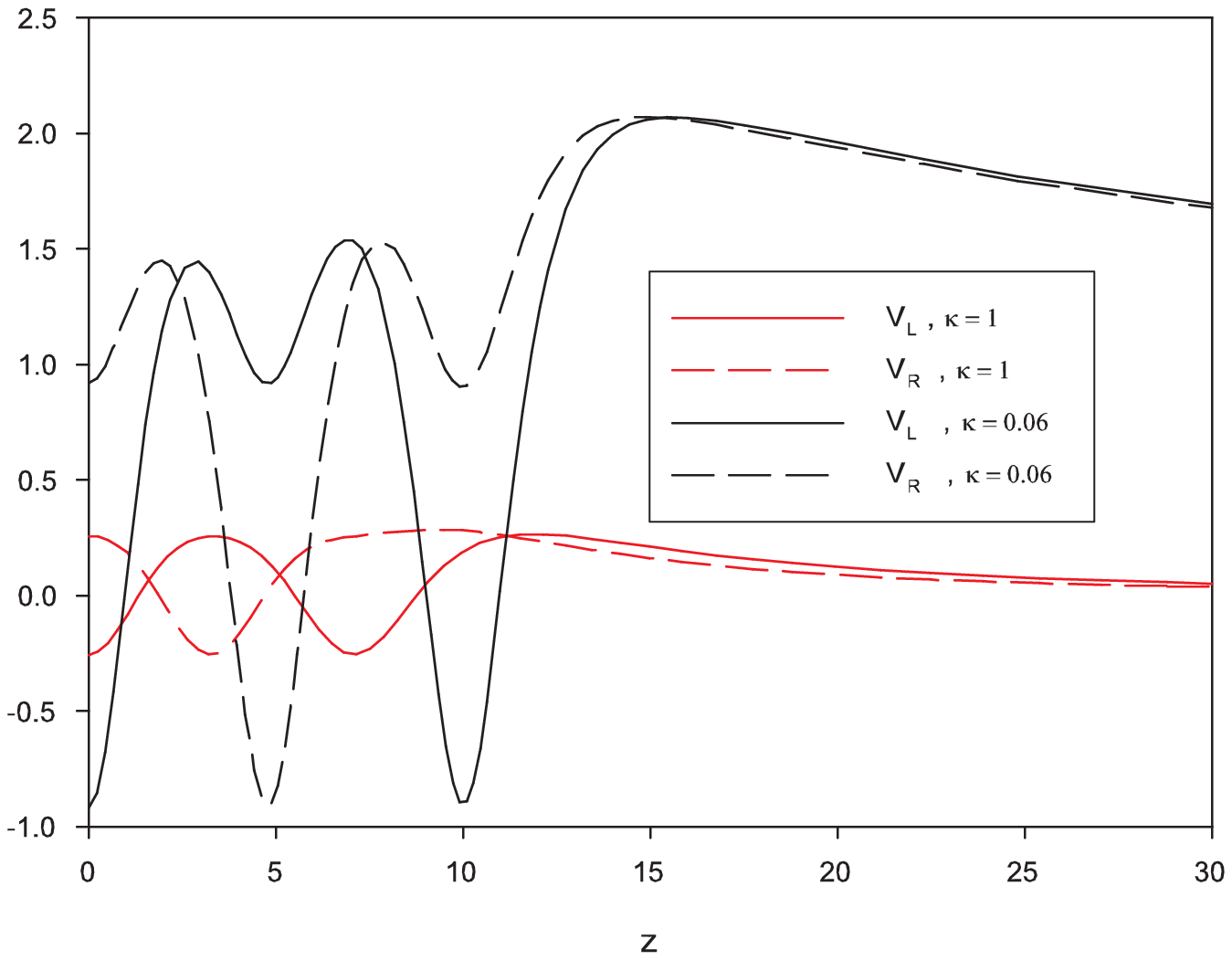}}
    \end{center}
   \caption{The effective potentials for the first (left) and second (right) excited branes
for $\eta = 1$ and two values of the : $\kappa$ parameter: 
$\kappa = 1, 0.003$ for the left and $\kappa=1, 0.006$ for the right graph. }
\label{effective_potential} 
 \end{figure}

Taking the next excitation, reveals that the corresponding effective potentials
develop more and more local minima in the bulk, as shown on Fig.\ref{effective_potential}.

\section{Stability}
\subsection{Setup}
We consider the following ansatz for the metric and matter field perturbations:
\be
\label{apertu}
ds^2 = e^{-2(A+\epsilon A_1(t,y))}\eta_{\mu\nu}dx^\mu dx^\nu + e^{-\epsilon B_1(t,y)}dy^2 + \epsilon C_1(t,y)dt dy \ ,
\  f(t,y) = f_0 + \epsilon f_1(t,y),
\ee
where $\epsilon$ is a small bookkeeping parameter and $A, f_0$ are solutions to the background equations given in the previous section (note that, for convenience, the function  $f$ of the previous section is here renamed $f_0$). 
The form (\ref{apertu}) is the most general ansatz compatible with the symmetries of the background spacetime. The stability issue of thick branes in five dimensions has been adressed in \cite{Kobayashi:2001jd} and we basically follow the same approach. 

Assuming, as usual,  the fields to depend on time through a factor $e^{i \omega t}$, 
the nontrivial gravity perturbation equations reduce to 
\bea
&&3 A' \left(8 A_1'-B_1'\right)+B_1 \left(12 A'^2-\kappa  f_0'^2\right)
+2 \left(-3 A_1''+\kappa  f_0' f_1'+\kappa  f_1 V'(f_0)\right)=0,\nonumber\\
&&-24 A' A_1'-12 B_1 A'^2+3 i \omega  e^{2 A} C_1 A'
+6 \omega ^2 e^{2 A} A_1+\kappa    B_1 f_0'^2+2 \kappa  f' f_1'-2 \kappa  f_1 V'(f_0)=0,\nonumber\\
&&3 B_1 A'+6 A_1'-2 \kappa  f_1 f_0'=0.
\eea
The perturbed matter equation leads to
\be
-2 \left(-4 A' f_1'+B_1 V'(f_0)+f_1''\right)+f_0' \left(-i \omega  e^{2 A} C_1+8 A_1'-B_1'\right)
+2  f_1 \left(V''(f)-\omega ^2 e^{2 A}\right)=0
\ee

The gravitational equations are not  independent~: solving 
the second and the third for $B_1,C_1$ and  reconstructing $C_1',B_1'$ 
leads to an identity for the first equation.
Knowing the expressions for $B_1,C_1,B_1',C_1'$ and further defining 
\be
F:=f_1-\frac{A_1 f_0'}{A'},
\label{def_F}
\ee
leads to the following master equation for the perturbation $F(y)$~:
\be
F'' - 4A' F' +\left( \frac{4 \kappa  f_0' V'(f_0)}{3 A'(y)}-\frac{2 \kappa ^2 f_0'^4}{9 A'^2}+\omega ^2 e^{2 A}+\frac{8}{3} \kappa  f_0'^2-V''(f_0) \right) F=0.
\label{meq}
\ee

Using the following change of function and change of variable:
\be
F \rightarrow e^{\frac{3}{2}A} F,\ \ dy = e^{-A} dr,
\ee
the equation (\ref{meq})  can be set 
in the form of a Schrodinger equation~: 
\bea
&&-\frac{d^2}{dr^2} \tilde F +V_p \tilde F = \omega^2 \tilde F,\nonumber\\
&&V_p=\frac{2 \kappa ^2 \tilde f_0'^4}{9\tilde A'^2}-\frac{4 \kappa  e^{-2 \tilde A} \tilde f_0' V'(\tilde f)}{3 \tilde A'}+\frac{15}{4} \tilde A'^2-\frac{19}{6} \kappa  \tilde f_0'^2+e^{-2\tilde A} V''(\tilde f_0),
\eea
where we defined $\tilde f_0(r) = f_0(y(r)),\ \tilde A(r)=A(y(r)),\ \tilde F(r) = F(y(r)) $.

Note that the
zero mode associated with the translational invariance of the background fields leads to a trivial $F$. 
Indeed the values $A_1 = A',\ f_1=f_0'$ imply  $F=0$.

\subsection{Fundamental solution}
\begin{figure}
\centering 
\includegraphics[scale=.6]{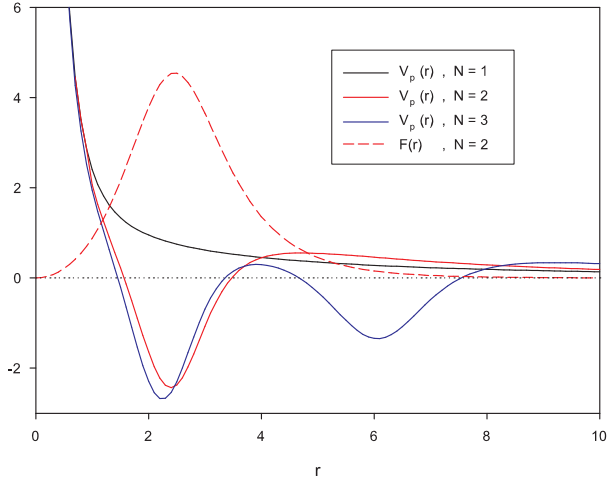}
\caption{Solid: potential of the perturbation master function for the solutions with one, two, three nodes.
Dashed: the first normalisable eigenmode of the two node solution.}
\label{pot_stabilite}
\end{figure}

For the range of the parameters considered, 
the fundamental solution is stable. Indeed, the potential is divergent close to the origin:
\be
\label{vzero}
\tilde A \approx A_{00},\ \tilde f\approx f_{01} r\ \Rightarrow V_p \approx \frac{2}{r},
\ee
and the asymptotic behaviour is 
\be
\label{vasymp}
\tilde A\approx \log r,\ \tilde f \approx \frac{\pi}{2}\ \Rightarrow V_p\approx \frac{4 V''\left(\frac{\pi }{2}\right)+15}{4 r^2}.
\ee
The profile of the corresponding potential is presented on Fig. \ref{pot_stabilite} by the solid black line.
The last term in the potential $V_p$ plays the role of the mass term, so it should be positive, 
leading to a positive fall-off of the eigenfunction. As a consequence, the potential is positive definite and there cannot be imaginary modes.


\subsection{Excited solutions}
The behaviour of the potential close to the origin (\ref{vzero}( and at infinity (\ref{vasymp}) 
are  the same as in the  case of the fundamental solution, 
but the potential  becomes negative in the intermediate region.
The potential corresponding to the  $N=2$ and $N=3$ solutions are represented on Fig. \ref{pot_stabilite}
by the red and blue lines respectively. 
The occurence of valleys where the potential is negative leads to 
bound states which possibly  reveals the existence of instable modes if  $\omega^2<0$. 
Indeed, we find an unstable mode for the generic solutions  with one mode.  For $\kappa = 1.0$ we find
 $\omega^2 \approx -0.695$. The profile of this eigen-mode (the normalisation is arbitrary) is supplemented 
 on figure \ref{pot_stabilite} by the dashed, red line.
 We further checked that the next bound state is indeed stable; this suggests that the brane 
 characterized by $N$ nodes of the scalar field has indeed $N-1$ unstable modes. 
 This speculation was, however, not tested further.

\section{Conclusions}
The purpose of this note is to exhibit new families of braneworld associated to five-dimensional
gravity
supplemented by a scalar sector admitting kink solutions.
The family of solutions is organized like
a tower of excited states of a Schrodinger equation 
(the bulk cosmological constant playing a role of spectral parameter, 
taking discrete values) although the underlying equations are highly non-linear. 
The types of braneworlds obtained in the paper have, to our knowledge, not been constructed
yet and present some potentially interesting features for the localization of both species
of chiral fermions. 

Often, when non linear equations admit various branches of solutions, 
the higher excitation are less stable than the main solution. 
It seems that this feature is obeyed in the present case~: 
using a particular channel of the perturbed equations
we checked that a typical solution with no nodes is stable and that the corresponding solution with one node is not. However, it could be expected that these new types of solutions is relevant in the mechanism of brane formations. Assuming brane formation in a dynamical process, possible unstable higher energy might be reached. The latter should then decay to the fundamental low energy configuration. 
These possibilities need further investigations.

\acknowledgments
T.D. acknowledges portuguese FCT and CERN for financial support through projects  PTDC/FIS/098032/2008 and CERN/FP/123593/2011.

\end{document}